\documentclass[10pt,letterpaper]{article}
\usepackage{opex3}

\begin{document}

\title{Waveguide grating mirror in a fully suspended 10\,meter Fabry-Perot cavity}

\author{Daniel Friedrich$^{1,*}$, Bryan W. Barr$^2$, Frank Br\"uckner$^3$, Stefan Hild$^2$ John Nelson$^2$, John Macarthur$^2$, Michael V. Plissi$^2$, Matthew P. Edgar$^2$, Sabina H. Huttner$^2$, Borja Sorazu$^2$, Stefanie Kroker$^3$, Michael Britzger$^1$, Ernst-Bernhard Kley$^3$, Karsten Danzmann$^1$, Andreas T\"unnermann$^3$, Ken A. Strain$^2$, and Roman Schnabel$^1$}

\address{$^1$ Max-Planck-Institut f\"ur Gravitationsphysik (Albert-Einstein-Institut) and Institut f\"ur Gravitationsphysik, Leibniz Universit\"at Hannover, Callinstrasse 38, 30167 Hannover, Germany\\
$^2$ SUPA, School of Physics and Astronomy, University of Glasgow, Glasgow G12 8QQ, UK\\
$^3$ Institut f\"ur Angewandte Physik, Friedrich-Schiller-Universit\"at Jena, Max-Wien-Platz 1, 07743 Jena, Germany}

\email{Daniel.Friedrich@aei.mpg.de} 



\begin{abstract}
We report on the first demonstration of a fully suspended $10\,$m Fabry-Perot cavity incorporating a waveguide grating as the coupling mirror. 
The cavity was kept on resonance by reading out the length fluctuations via the Pound-Drever-Hall method and employing feedback to the laser frequency. From the achieved finesse of $790$ the grating reflectivity was determined to exceed $99.2\,\%$ at the laser wavelength of $1064\,$nm, which is in good agreement with rigorous simulations. Our waveguide grating design was based on tantala and fused silica and included a $\approx 20\,$nm thin etch stop layer made of Al$_2$O$_3$ that allowed us to define the grating depth accurately during the fabrication process. Demonstrating stable operation of a waveguide grating featuring high reflectivity in a suspended low-noise cavity, our work paves the way for the potential application of waveguide gratings as mirrors in high-precision interferometry, for instance in future gravitational wave observatories.    
\end{abstract}



\section{Introduction}
Upcoming ground-based laser interferometric gravitational wave detectors  such as Advanced LIGO \cite{AdvLIGO}, Advanced Virgo \cite{AdvVirgo}, GEO-HF \cite{GeoHF} and LCGT \cite{LCGT} are expected to be limited by thermal noise in their most sensitive frequency band around $100\,$Hz. Currently, strategies for detectors beyond the 2nd generation, such as the 'Einstein Telescope' \cite{ET,ET2}, are being developed aiming for a ten times better sensitivity. Following Levin's approach \cite{Levin98,Harry02} mechanical dissipation located at a mirror front surface contributes more than the same dissipation within the substrate. The dominant contribution to thermal noise in the detection band of advanced detectors arises from multilayer coatings made of tantala and silica, used for highly reflective test mass mirrors at a laser wavelength of $1064\,$nm \cite{Harry02,Nawrodt11}. Hence, a reduction of the amount of mechanically lossy coating material is expected to improve the coating Brownian thermal noise of a test mass mirror. One approach being considered is based on resonant waveguide grating structures as a substitute for commonly used multilayer coatings. In particular, broadband waveguide grating structures under normal incidence have been proposed as an alternative concept for test mass mirrors, providing high reflectivity while having significantly less \cite{Bunk06}, or perhaps even no \cite{Brue08}, additional coating material. Both concepts have been successfully realized and tested in bench-top cavity experiments where the highest reflectivities seen so far are $99.08\,\%$ at $1064\,$nm \cite{Brue09} and $99.79\,\%$ at $1550\,$nm \cite{Brue10}. 

Here, we report on the next step towards the implementation of waveguide grating mirrors in large-scale gravitational wave detectors, namely their usage as a coupling mirror in a fully suspended low-noise prototype environment. Therefore, the Glasgow prototype facility \cite{Edgar,Barr,Huttner} was commissioned to incorporate a custom made waveguide grating mirror in a $10\,$m linear Fabry-Perot cavity. We could demonstrate stable operation of the waveguide cavity with a measured finesse of about $790$. The waveguide grating design was based on a tantala layer and silica substrate having an additional etch stop layer of Al$_2$O$_3$, which allowed us to define the grating depth accurately in the fabrication process. Its reflectivity was determined to be $\geq 99.2\,\%$, which is the highest reflectivity of such a device reported at a laser wavelength of $1064\,$nm so far.

\section{Waveguide grating mirrors under normal incidence}
\begin{figure}[htbp]
\begin{center}
\includegraphics[width=12cm]{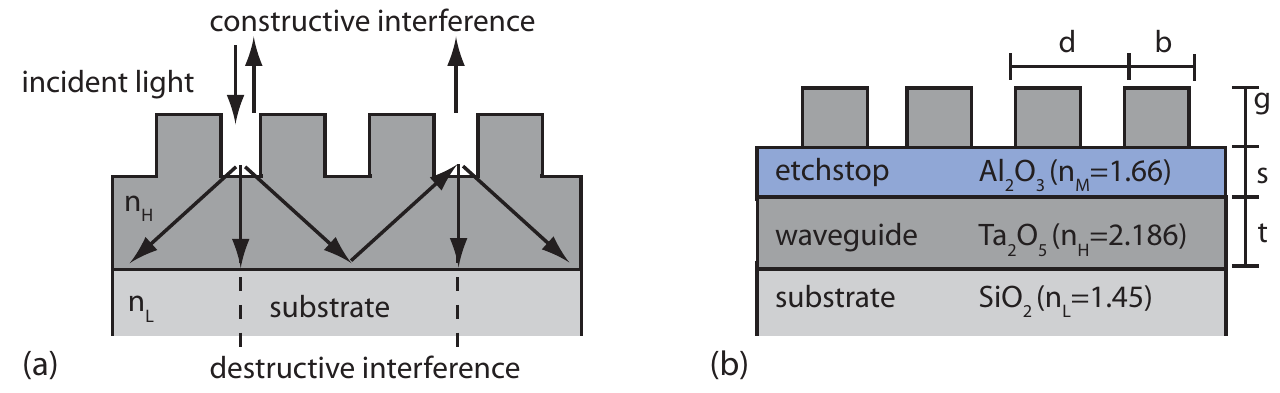}
\caption{(a) Principle of a waveguide grating mirror under normal incidence in a ray picture. The incoupled first order diffracted rays are guided via total internal reflection at the substrate with index of refraction $n_\mathrm{L}<n_\mathrm{H}$. The outcoupling at the grating structure is a constructive interference if all grating parameter are designed properly, hence, providing $100\,\%$ reflectivity. (b) Equivalent waveguide grating architecture realized in this work, including an Al$_2$O$_3$ layer as etch stop in order to define the grating depth in the fabrication process.} \label{fig:architecture}
\end{center}
\end{figure}

\begin{figure}[htbp]
\begin{center}
\includegraphics[width=12cm]{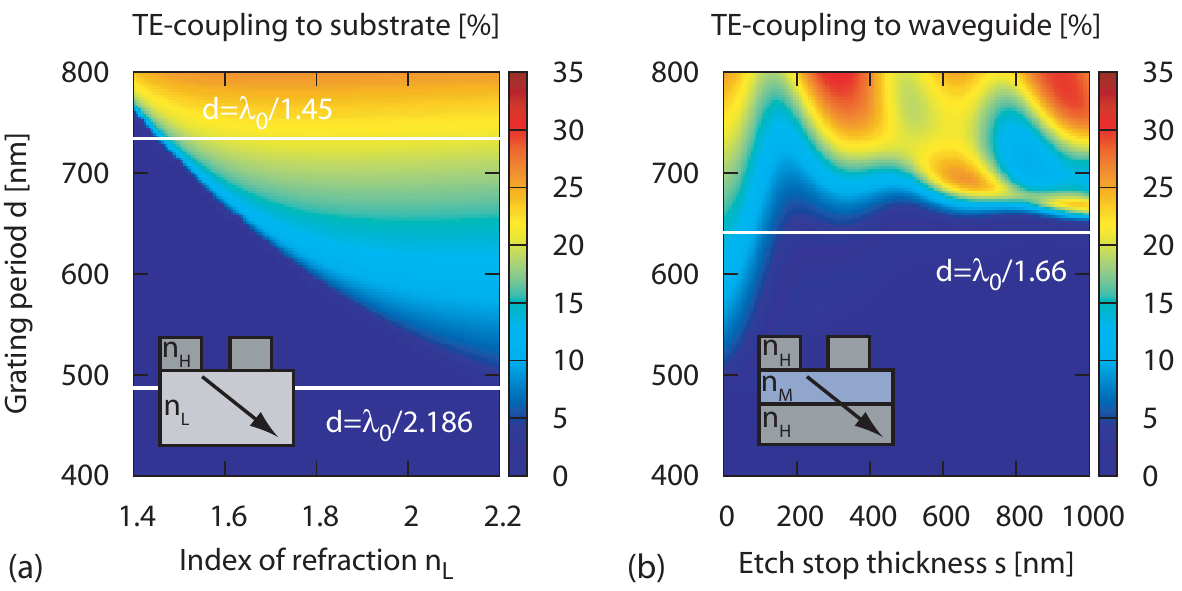}
\caption{(a) First order diffraction efficiency from a tantala grating ($n_\mathrm{H}=2.186$, $g=390\,$nm, $f=b/d=0.38$, TE-polarization) into a material with varying index of refraction $n_\mathrm{L}$, illustrating the range of the grating period that allows for resonant excitation as predicted by Eq.~(\ref{eq:range}). (b) If an etch stop layer with index of refraction  $n_\mathrm{M}=1.66$ is implemented, coupling into the waveguide occurs for $d\geq\lambda_\mathrm{0}/n_\mathrm{M}$ independent of its thickness $s$.} \label{fig:gratingperiod}
\end{center}
\end{figure}

A nanostructured waveguide layer can provide resonant excitation of an incoupling light field, which results in anomalies of its reflection coefficient  based on leaky waveguide modes as first found by Hessel and Oliner \cite{Hess65} and later experimentally confirmed for a first and zeroth diffraction order \cite{Mash84,Golu85,Mash85}. This property is, for instance, widely investigated for use in narrowband filter applications \cite{Magn92}. However, broadband designs under normal incidence as investigated in \cite{Bunk06,Brue08,Mate04,Brue09b} are favourable for the purpose of highly reflective surface mirrors in high-precision interferometric experiments, since they are less sensitive to parameter deviations. The basic principle of a waveguide grating mirror under normal incidence is shown in Fig.~\ref{fig:architecture}(a) using a ray picture \cite{Shar97}. 
It follows the idea of having a single nanostructured layer with high index of refraction $n_\mathrm{H}$ on a substrate having a lower index of refraction of $n_\mathrm{L}$. Incident light of wavelength $\lambda_\mathrm{0}$ can then be resonantly enhanced for subwavelength grating structures if the grating parameters, namely the grating period $d$, waveguide layer thickness $t$, grating depth $g$ and fill factor $f$ (ratio of ridge width $b$ and grating period $d$), are designed properly. From the grating equation one can derive a lower boundary for the grating period that permits the propagation of first diffraction orders in the waveguide layer and an upper boundary that allows guidance of light due to total internal reflection at the interface of high and low index of refraction materials \cite{Bunk06}. The same boundaries simultaneously prohibit higher diffraction orders in air ($>0$) and the waveguide layer ($>|\pm1|$). As a result, the grating period $d$ is restricted to a range of
\begin{equation}
487\,\mathrm{nm}=\lambda_\mathrm{0}/n_\mathrm{H} \leq d \leq \lambda_\mathrm{0}/n_\mathrm{L}=734\,\mathrm{nm} \label{eq:range}, 
\end{equation}
for tantala (Ta$_2$O$_5$) with $n_\mathrm{H}=2.186$ and fused silica (SiO$_2$) with $n_\mathrm{L}=1.45$ at a wavelength of $\lambda_\mathrm{0}=1064\,$nm. This restriction is illustrated in Fig.~\ref{fig:gratingperiod}(a) by means of Rigorous Coupled Wave Analysis (RCWA) calculations \cite{Moha81} for a tantala grating structure and TE-polarized light (electric field vector parallel to the grating ridges). The white lines mark the upper and lower boundary for the grating period with respect to the indices of refraction for the substrate and waveguide layer material, respectively.     

For the waveguide grating described in this article, we have implemented an additional thin layer of Al$_2$O$_3$ (see Fig.~\ref{fig:architecture}(b)), which was used as an etch stop (there is a high contrast of etching rates between Al$_2$O$_3$ and Ta$_2$O$_5$) to define a particular grating depth in the fabrication process. Its index of refraction of $n_\mathrm{M}=1.66$ is high enough to allow first order coupling to the waveguide layer if $d\geq\lambda_\mathrm{0}/n_\mathrm{M}$ holds for the grating period independent of the etch stop layer thickness $s$ (see Fig.~\ref{fig:gratingperiod}(b)). In particular for a thickness of only a few nanometers there is no significant influence on the coupling efficiency and consequently the waveguide grating's optical properties in comparison with the conventional design ($s \rightarrow 0\,$nm in Fig.~\ref{fig:gratingperiod}(b)).

\begin{figure}[htbp]
\begin{center}
\includegraphics[width=12cm]{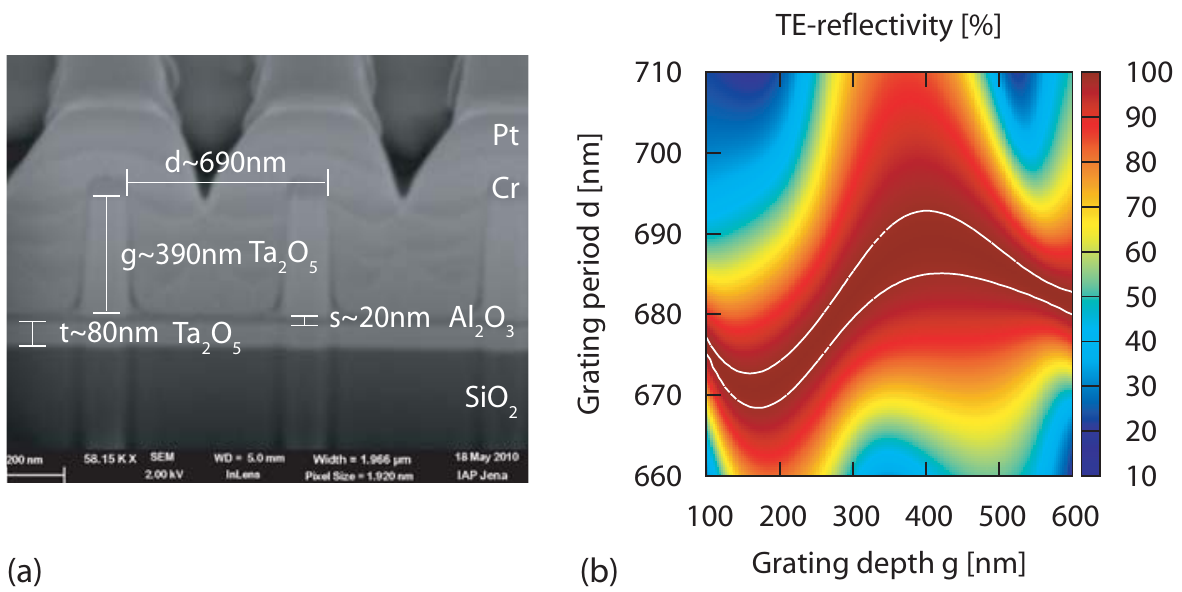}
\caption{(a) SEM image of a fabricated waveguide grating structure. Except of the fill factor all parameters are the same for the sample investigated here. (b) Calculated reflectivity (RCWA) of a waveguide grating under normal incidence for the parameters marked in the SEM image and a fill factor of $f=0.38$. The white lines mark an area of reflectivity $\geq 99\,\%$.} \label{fig:SEMdesign}
\end{center}
\end{figure}

The starting point for the fabrication was a standard $5\,$inch fused silica mask blank coated by a layer system of tantala ($t=80\,$nm), Al$_2$O$_3$ ($s=20\,$nm) and tantala ($g=390\,$nm). A top chromium (Cr) layer of $60\,$nm thickness was attached onto the layer system, serving as the mask during tantala etching. This mask was realized by spin coating the whole sample with an electron beam sensitive resist and applying electron beam lithography for an area of $(10\times15)\,$mm aiming at a grating period of $d=688\,$nm, which satisfies $\lambda_\mathrm{0}/n_\mathrm{M} \leq d \leq \lambda_\mathrm{0}/n_\mathrm{L}$, and a fill factor of about $0.38$. After resist development the chromium layer was structured by utilizing an Inductively Coupled Plasma (ICP) dry-etching process. Finally the binary chromium mask was transferred only into the upper tantala layer by means of an anisotropic Reactive Ion Beam Etching (RIBE) process supported by the high etching contrast between tantala and Al$_2$O$_3$. A scanning electron microscope (SEM) cross-sectional image of a fabricated structure from this process is shown in Fig.~\ref{fig:SEMdesign}(a), which is based on the same layer system and etching process but different fill factor as the one used in the experiment. 
The sample preparation for SEM characterization was done by using a focused ion beam (FIB) which requires covering the grating spot of interest by a platinum (Pt) layer (see Fig.~\ref{fig:SEMdesign}(a)). This image also shows the residual chromium mask, which was removed prior to sample application.

Corresponding to the values for the grating parameters given in Fig.~\ref{fig:SEMdesign}(a), RCWA was used to predict the reflectivity for TE-polarized light and a fill factor of $b/d=0.38$ (see Fig.~\ref{fig:SEMdesign}(b)). These results indicate that reflectivities higher $99\,\%$ are feasible, within parameter uncertainties of a few nanometers that are difficult to resolve on the basis of the SEM image. The most crucial parameter here is the tantala layer thickness $t$. A variation of this layer thickness by $\pm 3\,$nm leads to an absolute decrease in reflectivity of $\approx 1\,\%$, since the area of high reflectivity in Fig.~\ref{fig:SEMdesign}(b) shifts to larger grating periods with smaller waveguide layer thickness (not shown here).

We would like to note, that the etch stop layer design in principle offers the possibility to reduce the grating depth $g$ of an already fabricated grating sample without affecting the other parameters by means of an additional etching step, which has not been done yet for our investigated device. According to Fig.~\ref{fig:SEMdesign}(b) this could efficiently be used in conjunction with a slightly smaller grating period (or in case of a tantala layer thickness $t< 80\,$nm), which enables to cross the area of high reflectivity with decreasing grating depth.

\section{The 10 meter waveguide grating cavity}

\begin{figure}[htbp]
\begin{center}
\includegraphics[width=12cm]{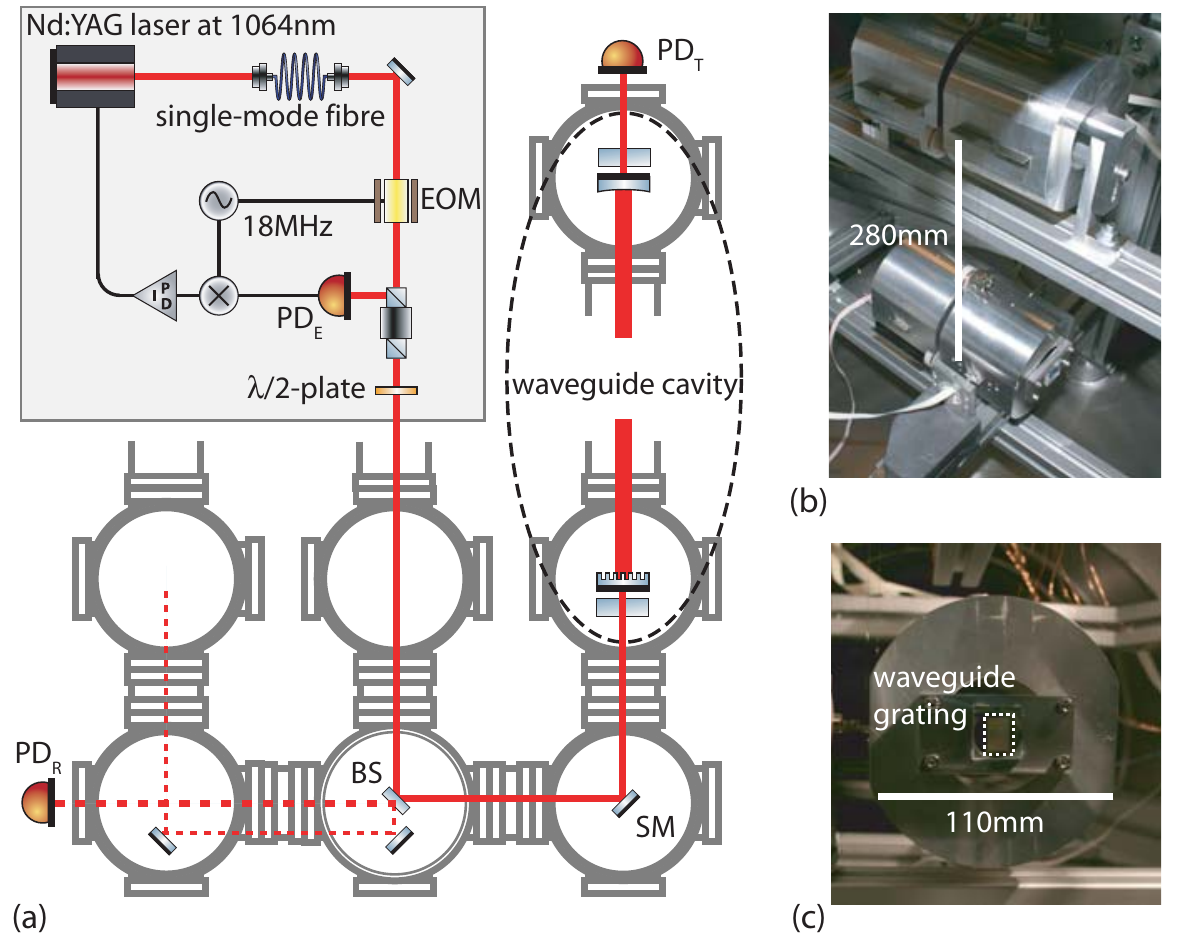}
\caption{(a) Schematic of the prototype facility including the laser bench, vacuum system and waveguide cavity having a length of $\approx 10\,$m. The cavity was stabilized using the Pound-Drever-Hall scheme also depicted at the laser bench.  (b) Intermediate and lower stage of the triple suspension system used for the cavity mirrors. Behind the main test mass a second triple suspension carries the so-called reaction mass, which is used to act on the main test mass. (c) Test mass with the waveguide grating mirror (area of $10\times 15\,$mm) attached.} \label{fig:JIF}
\end{center}
\end{figure}

Figure~\ref{fig:JIF}(a) shows a schematic overview of the section of the Glasgow 10m prototype used for the work described in this article. 
A Nd:YAG laser at a wavelength of $1064\,$nm is spatially filtered by a single-mode fiber before being injected to the vacuum system and guided to the waveguide cavity via a beam splitter (BS) and a steering mirror (SM) that both are realized as double suspensions. The cavity mirrors are suspended as triple pendulums (see Fig.~\ref{fig:JIF}(b)) based on the GEO\,600 suspension design \cite{Pliss00}. While the end mirror was a conventional multilayer mirror with a nominal power reflectivity of $|r_2|^2=0.9996\,$ the coupling to the cavity was realized with the waveguide grating mirror having an area of $(10\times15)\,$mm (see Fig.~\ref{fig:JIF}(c)). Their separation was $L_0=9.78\,$m, yielding a free spectral range of $FSR=c/(2L_0)=15.33\,$MHz with $c$ being the speed of light. The radius of curvature of the end mirror ($\approx 15\,$m) defines the beam waist diameter on the plane waveguide grating to be $3.1\,$mm, which is at least three times smaller than the grating. Therefore, we assume any power losses from beam clipping to be negligible. 

For all measurements the vacuum system had been evacuated to $\leq 10^{-4}\,$mbar in order to suppress acoustic noise and residual gas pressure noise. The cavity could be stabilized on resonance using the Pound-Drever-Hall scheme \cite{Drev83}. A phase modulation at $18\,$MHz was imprinted on to the incident light via an electro-optical-modulator (EOM), which is equivalent to sidebands at $\Omega=2.67\,$MHz from the first FSR. The reflected light was detected with the photodiode PD$_\mathrm{E}$ and electronically demodulated resulting in an bipolar error signal with zero crossing at the cavity resonance, which was fed back to the laser frequency. 

\begin{figure}[htbp]
\begin{center}
\includegraphics[width=12cm]{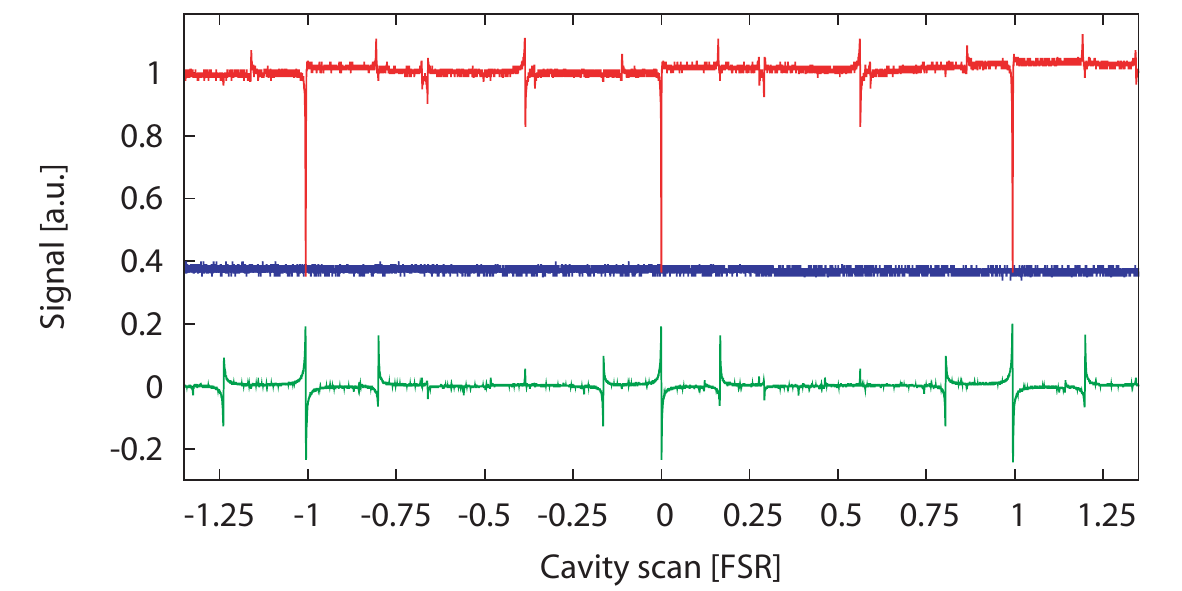}
\caption{Cavity scan via tuning of the laser frequency. Reflected power (red trace) and Pound-Drever-Hall error-signal (green trace) detected with the photodiodes  PD$_\mathrm{R}$ and PD$_\mathrm{E}$, respectively. The reflected signal for a stabilized cavity (blue trace) indicates a visibility of $\geq 0.6$.} \label{fig:FSR}
\end{center}
\end{figure}

In Fig.~\ref{fig:FSR} the reflected light power (red trace) and the error-signal (green trace) are shown, while the laser frequency was ramped. The reflected signal for the stabilized cavity (blue trace) shows a visibility of $\approx 0.6$, which is only a lower value due to power present in higher order modes and signal-sidebands.
 
\begin{figure}[htbp]
\begin{center}
\includegraphics[width=12cm]{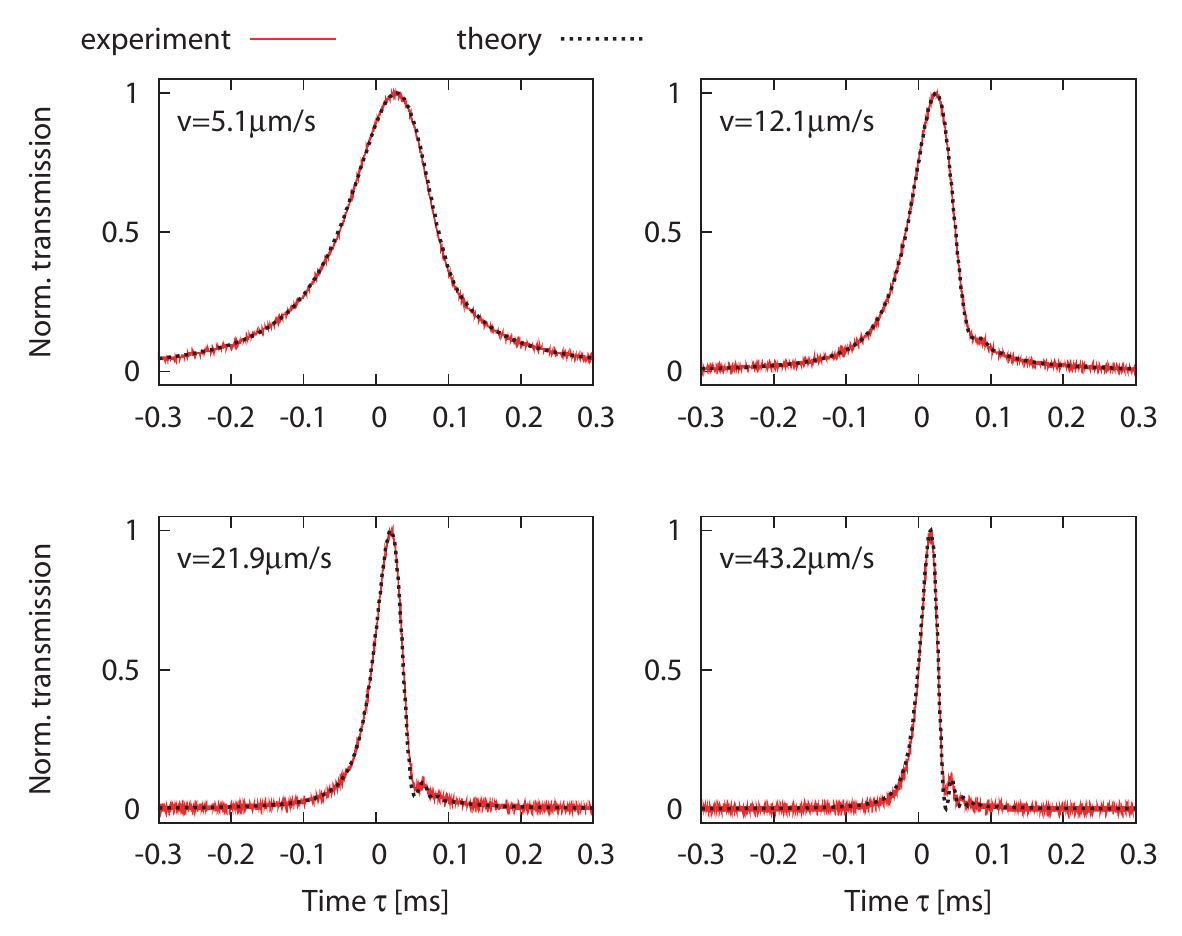}
\caption{Typical measurements of the transmitted light of a swept cavity resonance for different mirror velocities $v$ (red trace). Theoretical results for the transmitted power $|a_\mathrm{T}|^2$ are based on Eq.~(\ref{eq:trans}) with $r_1 r_2=0.996$ (black line), which corresponds to a cavity finesse of 790. Hence, a lower boundary for the waveguide grating power reflectivity is $|r_1|^2\geq 0.992$.} \label{fig:ring}
\end{center}
\end{figure}

The cavity finesse was already high enough to exhibit the dynamical effect of ringing for a sweep through resonance as shown in Fig.~\ref{fig:ring} for different mirror velocities $v$. 
Therefore, the cavity end mirror was pushed longitudinally via its reaction mass (magnet-coil-actuators). The end mirror's velocity with respect to the coupling mirror was determined via 
\begin{eqnarray}
v=\frac{2\Omega}{FSR}\frac{\lambda}{2}\frac{1}{\Delta \tau}, \label{eq:v}
\end{eqnarray} 
where $\Omega$ is the sideband frequency at $2.67\,$MHz and $\Delta \tau$ is the separation in time of the two sideband signals around resonance.  

The theoretical model for the light field dynamics of a swept cavity used here (see e.g. \cite{Dang70}) is based on a constantly moving end mirror with a velocity $v \ll c$, giving a time dependent cavity length of $L(\tau)=L_0+v\tau$. The transmitted field $a_\mathrm{T}(\tau)$ consists of a number of partial beams $a_m$ that have undergone $2m+1$ cavity transits. Their sum is written as
\begin{eqnarray}
a_\mathrm{T}(\tau)=a_0 \sum_{m=1}^{\infty} t_1 t_2 \left( r_1 r_2 \right)^{m-1}\mathrm{exp}(i\phi_m(\tau)), \label{eq:trans}
\end{eqnarray}  
where $a_0$ is the incident light field and $r_1, t_1$ and $r_2, t_2$ denote the amplitude reflectivity and transmissivity of the coupling and end mirror, respectively.
The round-trip phase for the $m$-th partial field can be approximated to  
\begin{eqnarray}
\phi_m(\tau) \approx (2m-1)\frac{2\pi L(\tau)}{\lambda}- m(m-1)\frac{2L_0}{c}v\frac{2\pi}{\lambda}, \label{eq:phi}
\end{eqnarray}
with the assumption that the cavity length change with each round-trip is negligible for calculating the round-trip time $\approx 2L_0/c$. The first term in Eq.~(\ref{eq:phi}) describes a cavity where at each time $\tau$ an equilibrium for the intra-cavity field is reached thus giving the well-known airy peaks. The second term does account for additional phases due to the non-zero round-trip time of light with respect to the mirror motion.     

The theoretical results in Fig.~\ref{fig:ring} are based on a product of amplitude reflectivities of $r_1 r_2=0.996 \pm 0.0005$, which yields a cavity finesse of $790 \pm 100$. The error given here arises from the uncertainty of the cavity length of $\pm 0.1\,$m and measured time separation $\Delta \tau$ of $3\,\%$, which both determine the mirror velocity via Eq.~(\ref{eq:v}). If we assume a perfect end mirror $r_2=1$ and zero round-trip loss a lower value for the waveguide grating power reflectivity is $|r_1|^2\geq 0.992 \pm 0.001$. This value is in good agreement with the numerical prediction shown in Fig.~\ref{fig:SEMdesign}(b) ($d=690\,$nm, $g=390\,$nm) thus supporting the principle of the waveguide grating architecture investigated here.

\section{Conclusion}
We have demonstrated stable operation of a fully suspended $10\,$m cavity incorporating a waveguide grating as coupling mirror, which is a key step towards the application of waveguide mirrors as test masses of future gravitational wave detectors.
A cavity finesse of about $790$ was determined from the deformation of airy peaks (i.e. ringing effect) for a sweep over the cavity resonance. The corresponding waveguide grating reflectivity of $\geq 99.2\,\%$ is, to the best of our knowledge, the highest reflectivity ever realized for a resonant waveguide grating at a laser wavelength of $1064\,$nm. The result is in good agreement with rigorous simulations, thus supporting the principle of the investigated design, which in principle enables accurate control of the grating depth via further etching. Whether the optical and mechanical properties of an additional thin etch stop layer made of  Al$_2$O$_3$ can meet the strict requirements for future detector test masses need to be further investigated.   

In future we also plan to characterize in more detail the noise performance of the suspended waveguide mirror. A special focus will be put on investigating the level of any potential phase noise coupling from side motion \cite{Barr11} of the waveguide mirror.

\section*{Acknowledgments}
This work was supported by the Deutsche Forschungsgemeinschaft within the Sonderforschungsbereich TR7, the Excellence Cluster QUEST and by the University of Glasgow and the Science and Technology Facilities Council (STFC). We also like to acknowledge the support of the IMPRS on Gravitational Wave Astronomy and the European Commission under the Framework Programme 7 (Grant Agreement 211743).  
\end{document}